\documentclass[layout=twocolumn]{achemso}

\usepackage[T1]{fontenc} % Use modern font encodings

\usepackage{amsmath,amsfonts,amssymb,flafter,float,graphicx,latexsym,natbib,verbatim,xfrac}
\usepackage[english]{babel}
\usepackage[usenames]{color}
\usepackage[mathscr]{euscript}
\usepackage[colorlinks=true,linkcolor=blue,urlcolor=blue,citecolor=blue]{hyperref}

\author{E. David Guarin Castro}
\affiliation{Departamento de Física, Universidade Federal de São Carlos, 13565-905 São Carlos, SP, Brazil}
\email{ed.guarin.castro@gmail.com}

\author{A. Pfenning}
\affiliation{Technische Physik, Physikalisches Institut and Röntgen Center for Complex Material Systems (RCCM), Universität Würzburg, Am Hubland, D-97074 Würzburg, Germany}

\author{F. Hartmann}
\affiliation{Technische Physik, Physikalisches Institut and Röntgen Center for Complex Material Systems (RCCM), Universität Würzburg, Am Hubland, D-97074 Würzburg, Germany}

\author{G. Knebl}
\affiliation{Technische Physik, Physikalisches Institut and Röntgen Center for Complex Material Systems (RCCM), Universität Würzburg, Am Hubland, D-97074 Würzburg, Germany}

\author{M. Daldin Teodoro}
\affiliation{Departamento de Física, Universidade Federal de São Carlos, 13565-905 São Carlos, SP, Brazil}

\author{Gilmar E. Marques}
\affiliation{Departamento de Física, Universidade Federal de São Carlos, 13565-905 São Carlos, SP, Brazil}

\author{S. Höfling}
\affiliation{Technische Physik, Physikalisches Institut and Röntgen Center for Complex Material Systems (RCCM), Universität Würzburg, Am Hubland, D-97074 Würzburg, Germany}

\author{G. Bastard}
\affiliation{Département de Physique, Ecole Normale Supérieure de Paris (ENS), Université PSL (Paris Sciences and Letters), F75005, Paris, France}

\author{V. Lopez-Richard}
\affiliation{Departamento de Física, Universidade Federal de São Carlos, 13565-905 São Carlos, SP, Brazil}

\title{Optical mapping of non-equilibrium charge carriers}

\begin{document}

\makeatletter
\setlength\acs@tocentry@height{4.45cm}
\setlength\acs@tocentry@width{8.25cm}
\makeatother

\begin{tocentry}
\includegraphics{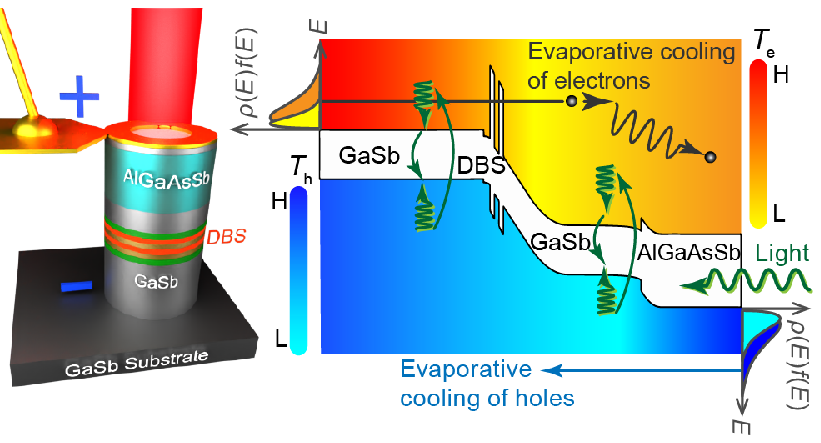}
\end{tocentry}

\begin{abstract}
We investigate the energy relaxation segmentation in a resonant tunneling heterostructures by assessing the optical and transport dynamics of non-equilibrium charge carriers. The electrical and optical properties are analyzed using electronic transport measurements combined with electro- and photoluminescence spectroscopies in continuous-wave mode. The radiative recombination is mainly governed by the creation of heavy holes \textit{via} impact ionization processes. Our results suggest hot electrons and holes populations form independent non-equilibrium systems that do not thermalize among them and with the lattice. Consequently, the carriers effective temperature changes independently at different regions of the heterostructure, with a population distribution for holes colder than for electrons.

\end{abstract}

Significant progress has been achieved in the electronic structure engineering of nanoscopic systems, allowing for a thorough characterization of the charge carrier dynamics and the correlation with the optical response. This is particularly crucial in semiconductor heterostructures like resonant tunneling diodes (RTDs)~\cite{growden2018,Ironside2019,Nie2019,Ahmadzadeh2020,Brown2020} where challenging questions still pervade the physics of carrier excitation, transport, relaxation, and recombination.~\cite{Lee2018,Lin_2020,Encomendero2020,Guarin2020,Cardozo2021} In particular, mapping the thermalization gradient along the transport path and how it is affected by external factors is still a relevant topic to be characterized and understood, and optical tools are well suited for that purpose.

The carrier thermalization affects nanodevices performance and their functionalities has attracted considerable attention in recent years.~\cite{Suchet2017,Xiao2018,Radevici2019,Sadi2020} It is also a source of fundamental problems related to energy transfer that can be tackled by studying the dynamics of hot-carriers and the concept of effective temperature.~\cite{Shah1985,Sze1997} This paper will shed light on these features by establishing the correlation between the transport and spectroscopic characterization of the device operation and subsequently explaining how the whole picture changes with an applied voltage and external illumination. With this approach, we have determined where the emitted light comes from, the role of the majority and minority carriers transport, and the energy relaxation segmentation along the heterostructure.

This analysis will offer clues on the relevant mechanisms that lead to the effective local temperature tuning, such as Joule heating, optical heating, or the evaporative cooling of hot-carriers. We also discuss the conditions when minority and majority non-equilibrium carriers are not able to mutually thermalize. The non-thermalization of different kinds of hot-carriers is an old prediction~\cite{Kalashnikov1977,Shah1985,Shah1992} grounded on the difference between carriers population and carriers effective masses. These differences control the balance between the efficiency of the carrier-carrier scattering process and the strength of the coupling of the carriers with the lattice that leads to potential energy losses.~\cite{Hopfel1986,Cui1988,Vandriel1992} Thus, the thermalization map may also discern temperature profiles according to the carrier type.

\section{Results and Discussion}

\begin{figure*}
	\includegraphics{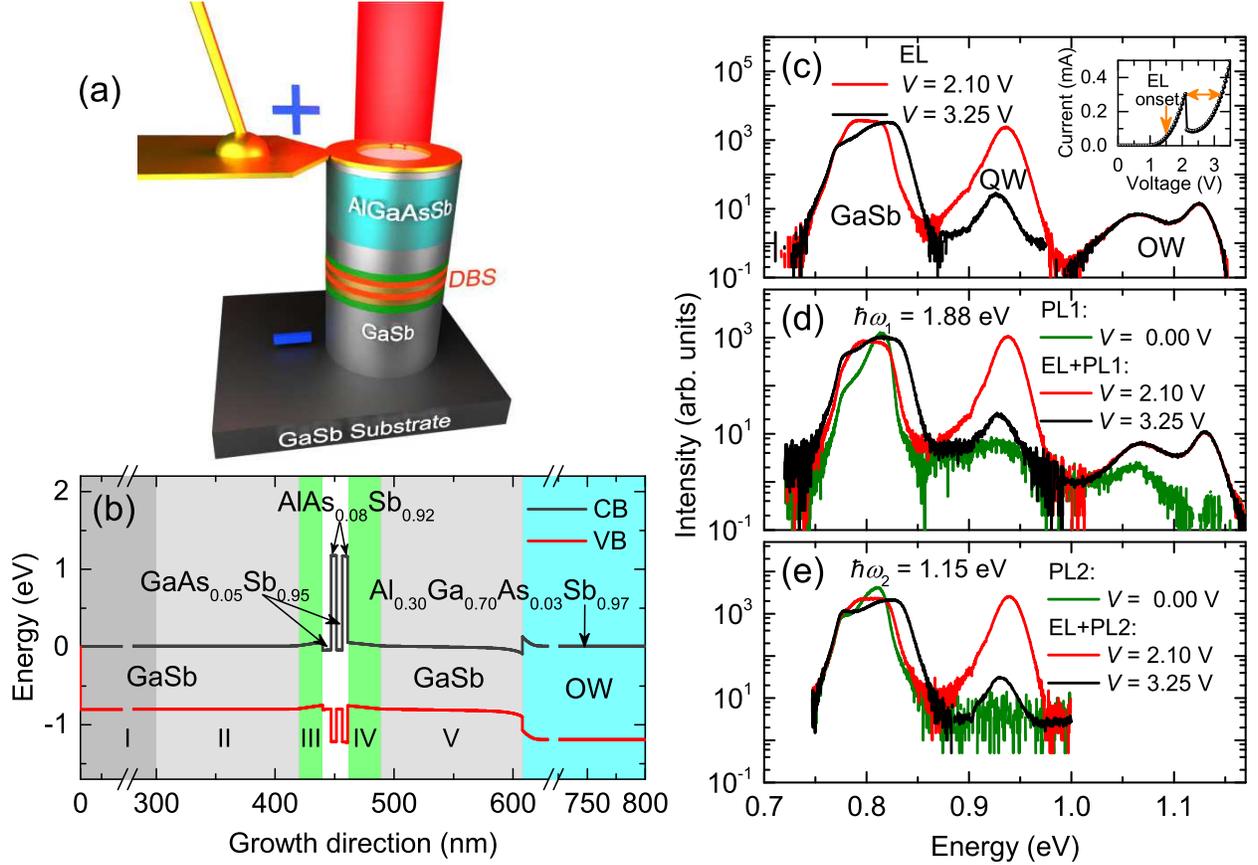}
	\caption{\label{Figure1} (a) Illustration of the RTD operation, indicating the applied-voltage configuration using in the experiments. (b) Simulated conduction band (CB) minimum (black line) and valence band (VB) maximum (red line). Roman numerals are used to differentiate the GaSb regions. (c) EL spectra at resonance ($V=2.10 \text{ V}$, red) and after resonance ($V=3.25 \text{ V}$, black) with the same current conditions as indicated by the $I(V)$-characteristic in the inset with a horizontal arrow. The EL onset is also indicated by a vertical arrow at $V=1.50 \text{ V}$. (d) EL+PL1 and (e) EL+PL2 spectra obtained for the same bias voltages as in (c), using the high ($\hbar\omega_1$) and low ($\hbar\omega_2$) excitation lasers, respectively. The pure PL spectra at $V=0.00 \text{ V}$ (green) are also shown.}
\end{figure*}

\textbf{Transport, generation, and recombination of carriers.} Figure~\ref{Figure1}(a) displays a schematic representation of the studied RTD, showing the applied-voltage configuration used during the experiments. The simulated conduction band (CB) minimum and valence band (VB) maximum along the RTD growth direction are presented in Figure~\ref{Figure1}(b) as black and red lines, respectively, for the zero-bias voltage condition.~\cite{Birner2007} Dark and light gray areas represent $n$-type GaSb layers with different doping levels. The $n$-type Al$_{0.30}$Ga$_{0.70}$As$_{0.03}$Sb$_{0.97}$ optical window (OW) is depicted as a cyan area. Green areas indicate undoped GaSb spacer layers, employed to avoid structural defects during the growing process of the double barrier structure (DBS),~\cite{Hartmann2012} which is depicted inside the white area. The DBS is formed by a pseudomorphically grown GaAs$_{0.05}$Sb$_{0.95}$ emitter prewell and quantum well (QW) and by two AlAs$_{0.08}$Sb$_{0.92}$ barriers. Roman numerals are employed to identify different GaSb regions.

Electroluminescence (EL) spectra are displayed in Figure~\ref{Figure1}~(c) for two bias voltages, $V=2.10\text{ V}$ and $V=3.25\text{ V}$, that correspond to the on-resonance and out-of-resonance transport conditions of the RTD. We should note that the current through the heterostructure is the same at these two bias voltages as pointed out in the current-voltage characteristic, $I(V)$, shown in the inset with a double-horizontal arrow. The two emission bands observed for highest energies, $1.12\text{ eV}$ and $1.06\text{ eV}$, correspond to band-to-band~\cite{Adachi1987} and donor-level transitions inside the Al$_{0.30}$Ga$_{0.70}$As$_{0.03}$Sb$_{0.97}$ optical window and are labeled as OW. The emission at $0.93\text{ eV}$, labeled as QW, corresponds to the interband recombination between the first confined levels in the double-barrier QW. Finally, the low-energy band ranging from $0.72\text{ eV}$ to $0.86\text{ eV}$ corresponds to combined emissions from bulk GaSb layers I, II, III, V.~\cite{Nakashima1981,Wu1993} The depletion of region IV by the applied forward voltage hampers there the radiative recombination process.

Under illumination and below the EL onset (also indicated in the inset of Figure~\ref{Figure1}~(c)), pure photoluminescence (PL) emission can be studied. Figures~\ref{Figure1}~(d) and (e) depict the PL1 and PL2 spectra (green lines) at zero-bias voltage, corresponding to the excitations with high ($\hbar\omega_1$) and low ($\hbar\omega_2$) laser energies, respectively. The absence of the most energetic emission at $1.12\text{ eV}$ in Figure~\ref{Figure1}~(d), which only appears for applied voltages above the EL onset, indicates a low-efficient photogeneration of electron-hole (e-h) pairs in the OW at this energy. The Al$_{x}$Ga$_{1-x}$As$_{y}$Sb$_{1-y}$ system presents a transition between direct- and indirect-energy-gap close to $x=0.30$~\cite{Adachi1987,dolginov1984} that may prevent the emission above $1.10\text{ eV}$. In turn, impurities in the OW due to the high doping profile can absorb light from the high-energy laser more efficiently and produce the less energetic OW emission observed at $1.06\text{ eV}$. For applied voltages above the EL onset, the PL and EL emissions unavoidably overlap, and the corresponding EL+PL spectra are shown in Figures~\ref{Figure1}~(d) and (e) as red and black lines.

\begin{figure}
    \includegraphics{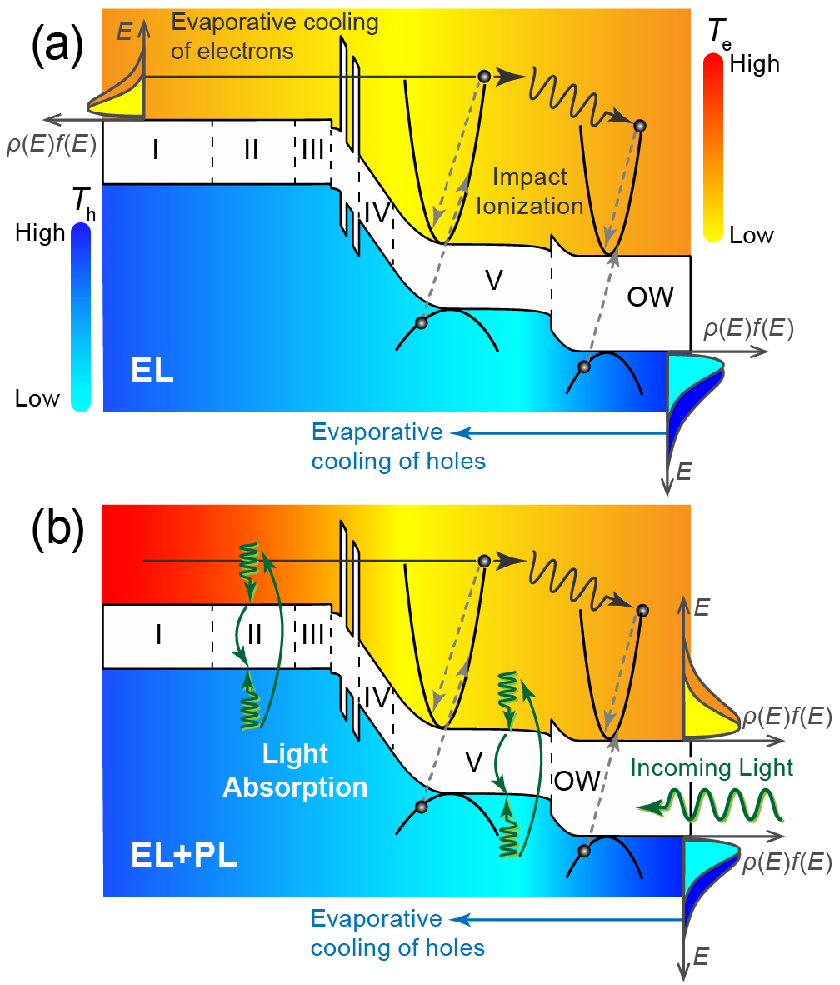}
    \caption{\label{Figure2} Energy band profiles and illustration of the charge carrier dynamics under an applied voltage. (a) In the absence of external illumination, holes produced after electrons impact ionization (gray dashed arrows) at region V and the OW are transported towards region I (straight blue arrow), accumulating in region IV, or recombining optically at intermediary regions. The carriers population distribution, $\rho(E)f(E)$ is represented at region I and the OW corresponding to the carriers effective temperature $T_i$ ($i=$ e,h). These distributions relax towards colder distributions due to the evaporative cooling of hot carriers.~\cite{Suchet2017} The difference between carriers effective masses allow electrons to reach higher temperatures (yellow gradient) than holes (blue gradient) (b) PL excitation gives additional energy to the carriers due to light absorption (green arrows), increasing $T_i$ and broadening $\rho(E)f(E)$ towards hotter distributions.}
\end{figure}

The bias voltage also affects the number of carriers available for optical recombination and the efficiency of this process in different parts of the heterostructure. Note in both Figures~\ref{Figure1}~(c) and (d) that the QW emission intensity changes drastically between $V=2.10\text{ V}$ and $V=3.25\text{ V}$ even though the current through the RTD remains the same, while the intensities of both the GaSb and OW are almost unchanged unveiling the contrast between resonant and non-resonant conditions. In turn, the high-energy side of the GaSb emission suffers a broadening towards higher energies after resonance, due to the elevation of the Fermi level at region III and in the prewell, caused by electrons accumulation.~\cite{Wu1993} 

The main processes contributing to the charge dynamics along the heterostructure have been represented in Figure~\ref{Figure2} within a sketch of the energy band profile under a forward bias voltage. The panel (a), in Figure~\ref{Figure2}, represents the transport of charge carriers and the holes generation through impact ionization in dark conditions that leads to the EL response. After tunneling through the DBS, a large amount of electrons coming from regions I-III can arrive ballistically at region V and the OW (straight gray arrow),~\cite{Sze1997,Yokoyama1988,Heiblum1990} where they can lose part of their energy by scattering processes (gray wavy arrow). However, the electrons' energy is still enough to induce impact ionization processes~\cite{White1991,Cardozo2018} producing holes, minority carriers that contribute to the optical recombination detected in the EL. These minority holes can also be transported to other regions inside the heterostructure (represented by a horizontal blue arrow at the bottom) under an applied voltage, even until region I. Under illumination, as depicted in Figure~\ref{Figure2}~(b), holes can be additionally generated by light absorption (green arrows), which occurs to a greater extent in the GaSb regions, where around $80\%$ of the incident light is absorbed according to calculations using the Lambert-Beer law~\cite{zimmermann2009} for the low-energy laser. The optical generation of e-h pairs under illumination leads to the PL below the EL threshold and contributes to the combined EL+PL above it. In the OW, the absorption of light is mainly due to impurity levels, as mentioned before, and only impact ionization can produce the holes necessary for recombination at the most energetic emission.

\begin{figure}
	\includegraphics{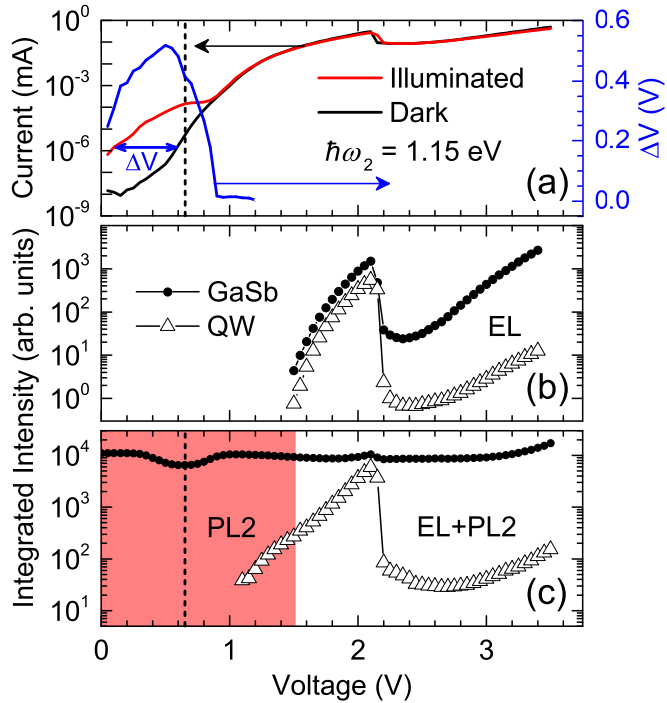}
	\caption{\label{Figure3} (a) $I(V)$-characteristic in the dark (black line) and under illumination measured with the low-energy laser (red line). At low bias voltages, the $I(V)$-characteristic presents a voltage shift ($\Delta V$, blue), with a maximum at $V=0.50 \text{ V}$. Integrated intensity as a function of the bias voltage for (b) EL and (c) EL+PL2 emissions, coming from QW (open triangles) and GaSb (solid circles) emissions. In (c), the red-shaded region represents the bias voltage range where just PL emission is observed. Under these conditions, a dip in the GaSb emission is detected at $V=0.65 \text{ V}$ (dashed line).}
\end{figure}

The $I(V)$-characteristics in the dark and under illumination are displayed in Figure~\ref{Figure3}~(a) on a semi-logarithmic scale. In both cases, resonant tunneling of majority electrons produces a peak current of $0.30\text{ mA}$ at the resonance voltage of $2.10\text{ V}$. For applied voltages between $0\text{ V}$ and $1.20\text{ V}$, the $I(V)$-characteristic is shifted to lower voltages during excitation with the low-energy laser. The photoinduced voltage shift, $\Delta V$, is produced by the accumulation of photogenerated holes in the depleted region IV and is proportional to their number.~\cite{Pfening2015,Pfening2018} The dependence of $\Delta V$ with the applied voltage is illustrated in Figure~\ref{Figure3}~(a) as a blue line, with a maximum of holes accumulation at $V=0.50\text{ V}$. This electrostatic effect will have a significant impact on the thermalization analysis unveiled below.

The correlation between the $I(V)$-characteristic in the absence of illumination and the EL emission is shown in Figure~\ref{Figure3}~(b), where the integrated EL intensities for QW (open triangles) and GaSb (solid circles) emissions are shown as functions of the applied voltage. Starting at $V=1.50\text{ V}$, both EL emissions increase with the resonant current, peaking at the resonance voltage with almost the same intensities. For higher bias voltages, the EL intensities decrease to minimum values at the valley voltage and then rise again due to impact ionization produced by non-resonant currents.~\cite{Cardozo2018} However, the QW intensity after resonance is noticeably lower than the GaSb intensity.

The contrasts between QW and GaSb emissions are more prominent during illumination, as illustrated by the PL integrated intensities as a function of the applied voltage in Figure~\ref{Figure3}~(c). The red-shaded region corresponds to the bias voltage range before the EL onset. Despite the QW and GaSb emissions peak at the resonance voltage with almost the same intensities, the absorption of light in the GaSb layers (especially in regions IV and V) creates photogenerated carriers, which enhance the radiative recombination for the whole bias voltage range. Therefore, the GaSb emission keeps an almost constant intensity, much higher than the QW emission, for out-resonance conditions. In contrast, the QW emission starting at $V=1.00\text{ V}$ maintains a clear correlation with the $I(V)$-characteristic.

A remarkable aspect presented in Figure~\ref{Figure3}~(c) is the dip in the GaSb intensity observed between $V=0.30\text{ V}$ and $V=0.90\text{ V}$ with a reduction of up to 42\% at $V=0.65\text{ V}$, regarding the intensity at zero-bias voltage. This minimum is correlated with the $\Delta V$ shift in Figure~\ref{Figure3}~(a) as indicated by the vertical dashed line and can be ascribed to the transport of photogenerated holes from region V and their accumulation in region IV. Below $V=0.50\text{ V}$, the growing accumulation of holes at region IV repels photogenerated holes at region III, lowering the probability of band-to-band optical recombination in this region. When the resonant tunneling of holes is triggered after $0.50\text{ V}$, not only the accumulation of holes in region IV is reduced, but also the minority carriers population shrinks at region V. Consequently, the GaSb intensity drops to its minimum value where only optical emission from regions I and II is expected. The intensity is restored when the accumulation of holes vanishes close to $V=0.90\text{ V}$ and photogenerated carriers can recombine again in regions III and V. This allows mapping the emissions from different GaSb regions in the sample by tuning the bias voltage range.

\textbf{Effective Carriers Temperature.} Considering that the optical emission rate ($Q$) is proportional to the joint density of states ($\rho_{\text{JDS}}$), $Q\propto \rho_{\text{JDS}}(\omega)$, where $\omega$ is the emitted photon frequency, the emission rate for uncoupled e-h pairs in stationary conditions, except for the QW, will be proportional to the probability distribution of each type of carriers in 3D,
\begin{multline} \label{Q1}
    Q \propto \sqrt{2m^{*\;3}_{\text{r}} (\hbar\omega-E_{\text{g}})}\\ f_{\mu_{\text{e}}^*}\left(\hbar\omega-E_{\text{g}},\frac{m^*_{\text{e}} T_{\text{e}}}{m^*_{\text{r}}}\right) 
    f_{\mu_{\text{h}}^*}\left(\hbar\omega-E_{\text{g}},\frac{m^*_{\text{h}} T_{\text{h}}}{m^*_{\text{r}}}\right)
\end{multline}
where $E_{\text{g}}$ is the bandgap energy, $m_i^*$ ($i=$ e,h) is the effective carrier mass, $m^*_{\text{r}}=m_{\text{e}}m_{\text{h}}/(m_{\text{e}}+m_{\text{h}})$ is the effective reduced mass, and $f_{\mu_i^*}$ is the carrier distribution function, characterized by an effective chemical potential, $\mu_i^*$: for electrons, $\mu_{\text{e}}^*=m^*_{\text{e}}(\mu_{\text{e}}-E_{\text{g}})/m^*_{\text{r}}$, and for holes, $\mu_{\text{h}}^*=m^*_{\text{h}}\mu_\text{h}/m^*_{\text{r}}$, with $\mu_i$ representing the chemical potential for each carrier population. Given the finite time for energy exchange between the carriers and the lattice, which is assumed in equilibrium with the environment at a temperature, $T_{\text{L}}$, the mean energy excess with respect to the equilibrium energy can be characterized by a carrier temperature, $T_i$. 
In the energy relaxation time approximation, the mean variation of energy over time is given by~\cite{Kalashnikov1977}
\begin{equation} \label{Ei}
    \left\langle \frac{dE_i}{dt} \right\rangle \approx \frac{\frac{2}{3} \left\langle E_i\right\rangle - k_{\text{B}} T_{\text{L}}}{\tau^\varepsilon_i}
\end{equation}
with $k_{\text{B}}$ as the Boltzmann constant, the energy relaxation time represented by $\tau^\varepsilon_i$, and the average carrier energy defined by $\left\langle E_i\right\rangle \equiv 3/2 k_{\text{B}} T_i$. Thus, $T_i$ is a parameter that depends on both external fields and scattering mechanisms. 

In the non-degenerate limit, eq. \ref{Q1} can be expressed as,
\begin{multline} \label{Q2}
    Q \propto \sqrt{2m^{*\;3}_{\text{r}} (\hbar\omega-E_{\text{g}})} \ \exp\left[\frac{-\hbar\omega}{k_{\text{B}} T^{\text{e-h}}_{\text{eff}}}\right] \\ 
    \exp\left[\frac{\left(\mu_{\text{e}}-E_{\text{g}}\right)+\frac{m^*_{\text{r}}}{m^*_{\text{e}}}E_{\text{g}}}{k_{\text{B}}\ T_{\text{e}}}+\frac{-\mu_{\text{h}}+\frac{m^*_{\text{r}}}{m^*_{\text{h}}}E_{\text{g}}}{k_{\text{B}}\ T_{\text{h}}}\right]
\end{multline}

The first exponential term in eq.~\ref{Q2}, which depends on the energy of the emitted photons ($\hbar\omega$), represents the Boltzmann's exponential decay ruled by the e-h effective temperature given by,
\begin{equation} \label{EqTeff}
    \frac{1}{T^{\text{e-h}}_{\text{eff}}}=\frac{m^*_{\text{r}}}{m^*_{\text{e}} T_{\text{e}}}+\frac{m^*_{\text{r}}}{m^*_{\text{h}} T_{\text{h}}}
\end{equation}

An analogous energy decay can be obtained for 2D e-h pairs. The forthcoming discussion will focus on the characterization of $T^{\text{e-h}}_{\text{eff}}$. 

\begin{figure*}
	\includegraphics{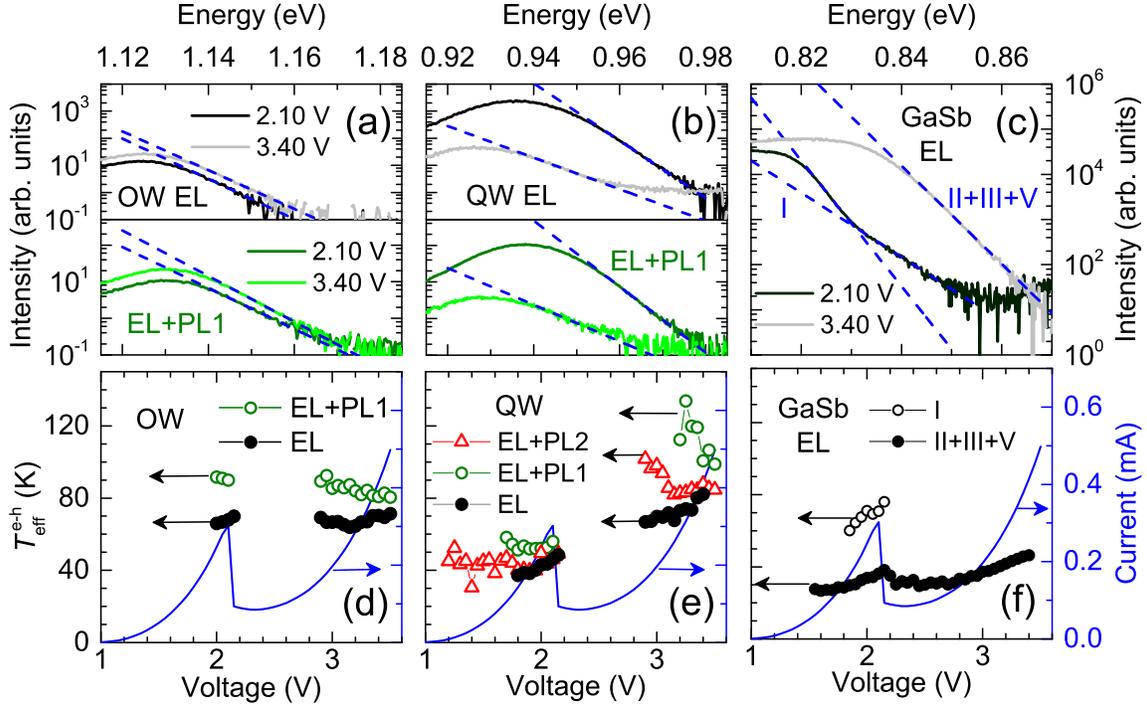}
	\caption{\label{Figure4} High-energy tails in the EL (dark lines) and EL+PL1 (green lines) spectra of the (a) OW, (b) QW, and (c) GaSb layers. The GaSb PL spectra is presented in Figure~\ref{Figure5}. The tails were fitted using Boltzmann functions (blue dashed lines). The spectra at resonance ($V=2.10\text{ V}$) and after resonance ($V=3.40 \text{ V}$) are displayed for comparison. Voltage dependence of $T^{\text{e-h}}_{\text{eff}}$ in the (d) OW, (e) QW, and (f) GaSb layers. $T^{\text{e-h}}_{\text{eff}}$ was obtained from EL (black dots), EL+PL1 (green dots), and EL+PL2 (red triangles) spectra. The $I(V)$-characteristic (blue solid line) is presented as reference. In (d) and (e), the low intensity of the spectra at the valley current condition impedes the extraction of a reliable $T^{\text{e-h}}_{\text{eff}}$.}
\end{figure*}

Figures~\ref{Figure4}~(a) and (b) show the high-energy tails of the OW and QW emissions, respectively, obtained from EL (top panels) and EL+PL1 (bottom panels) spectra at $V=2.10\text{ V}$ (dark lines) and $V=3.40\text{ V}$ (light lines). The high-energy tails for the same bias voltages obtained from the GaSb EL emission are also shown in Figure~\ref{Figure4}~(c). Dashed blue lines represent the corresponding Boltzmann’s distribution functions, $f\propto \exp\left[-\hbar\omega/\left(k_{\text{B}} T^{\text{e-h}}_{\text{eff}}\right)\right]$. Note that, albeit the QW and OW emissions present monoexponential high-energy tails for both EL and EL+PL1 spectra, the high-energy side of the GaSb EL emission exhibits a biexponential trend at the resonance condition with two spectral tails: (i) a high-intensity low-$T^{\text{e-h}}_{\text{eff}}$ tail associated to band-to-band optical recombination of carriers in regions II, III, and V and (ii), a low-intensity high-$T^{\text{e-h}}_{\text{eff}}$ tail coming from optical recombination of carriers in region I. The latter can be produced by the high doping profile,~\cite{Wu1993} which can increase the effective temperature, as reported earlier by Makiyama, \textit{et al.}~\cite{Makiyama1988}

The e-h pair effective temperatures obtained from the Boltzmann’s distribution functions for each emission band are plotted in Figures~\ref{Figure4}~(d)-(f) for EL (solid and open black circles), PL1 (open green circles), and PL2 (open red triangles) emissions. The $I(V)$-characteristic has been added as a reference. When the current is low at the valley, the intensity of the emitted light weakens and prevents the extraction of reliable values of $T^{\text{e-h}}_{\text{eff}}$. Note that $T^{\text{e-h}}_{\text{eff}}$ is always higher than the lattice temperature, $T_{\text{L}}=4\text{ K}$, indicating the presence of hot carriers along the whole heterostructure. 

According to Figure~\ref{Figure4}~(d), $T^{\text{e-h}}_{\text{eff}}$ remains constant at the OW during EL, with a mean value of $68 \text{ K}$. However, under illumination, optical heating increases $T^{\text{e-h}}_{\text{eff}}$, due to excitation of a large fraction of carriers to high energetic levels, after absorbing the excess energy given by the incoming photons.~\cite{Vandriel1992,Meyer1980} Then, $T^{\text{e-h}}_{\text{eff}}$ rises up to $90 \text{ K}$ before resonance, but for $V>2.90\text{ V}$, $T^{\text{e-h}}_{\text{eff}}$ drops to $80 \text{ K}$ suggesting a carriers cooling process, that can be ascribed to the evaporation from the OW of highly energetic holes.~\cite{Suchet2017}
The reason why this temperature drop can be only observed under illumination will be explained in the next section. 
In the case of the QW emission, the $T^{\text{e-h}}_{\text{eff}}$ increases with the applied voltage from $\sim37 \text{ K}$ before resonance, up to $134 \text{ K}$ after resonance, depending on the type of excitation as displayed in Figure~\ref{Figure4}~(e). The highest temperatures in the QW are achieved during the EL+PL1 emission and a cooling process similar to the cooling in the OW is also observed under illumination above $V=2.90\text{ V}$.

The cooling observed in both the OW and the QW is produced by evaporation of highly energetic carriers when they are extracted from a specific region, driven by the applied voltage,~\cite{Suchet2017,Ketterle1996} as depicted in Figures~\ref{Figure2}~(a) and (b) for electrons (horizontal black arrow) and holes (horizontal blue arrow). In stationary conditions, the population distribution can be characterized by $\rho(E)f(E)$, with $\rho(E)$ the density of states and $f(E)$ the distribution function characterized by $T^{\text{e-h}}_{\text{eff}}$. When the most energetic carriers are removed, the local state relaxes towards a ``colder'' distribution reducing the local carriers temperature, as exemplified in Figure~\ref{Figure2}~(a) by the schematic population distributions in region I for electrons and in the OW for holes.

Figure~\ref{Figure4}~(f) shows that the effective temperatures associated with the spectral tail coming from region I can only be resolved for bias voltages between $1.85 \text{ V}$ and $2.15 \text{ V}$. Note in Figure~\ref{Figure4}~(c) that after resonance, the emissions from regions II, III, and V broaden towards higher energies, masking the spectral tail from region I. The extracted temperatures from both tails also increases with the applied voltage ranging from $27 \text{ K}$ up to $46 \text{ K}$ for the II+III+V-tail, and from $60 \text{ K}$ up to $76 \text{ K}$ for the I-tail, as plotted in Figure~\ref{Figure4}~(f). The overall rise in the carriers effective temperature by increasing the applied voltage is expected due to Joule heating under high electric fields during their drift towards the outermost regions of the heterostructure, as illustrated by dark colors at the CB and VB in Figures~\ref{Figure2}~(a) and (b).~\cite{Heinrich1971,Hopfel1986,Sze1997,Das2015}

\begin{figure}
	\includegraphics[width=\columnwidth]{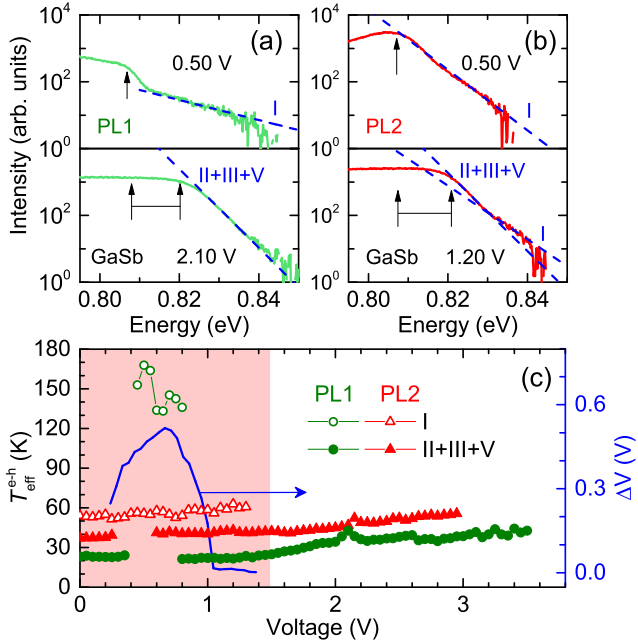}
	\caption{\label{Figure5} High-energy tails in the (a) PL1 (green lines) and (b) PL2 (red lines) GaSb spectra for two different bias voltages. In both cases, the II+III+V emission is shrunk at $V=0.5 \text{ V}$, while the emission from region I prevails. For other bias voltages, the II+III+V-tail is observed again due to the GaSb spectra broadening (black arrows). (c) $T^{\text{e-h}}_{\text{eff}}$ versus applied voltage obtained from the spectral tails of the PL1 (green dots) and PL2 (red triangles) spectra. The voltage shift ($\Delta V$, solid blue line) is shown as a reference inside the red-shaded voltage region where only PL emission is observed.}
\end{figure}

Now, let us explore in detail the GaSb spectra under illumination, as shown in Figure~\ref{Figure5}. Panels (a) and (b) correspond to the PL1 and PL2 high-energy side spectra. Top and bottom panels display the spectral tails coming from regions I and II+III+V, at low and high voltages, respectively. For both PL1 and PL2 emissions, a narrowing in the high-energy side of the GaSb spectra is evident at $V=0.50 \text{ V}$ when compared to higher bias voltages. The spectral narrowing results from the intensity loss of the emissions contributing to the II+III+V-tail, with energies higher than $807 \text{ meV}$ (black arrows). As already discussed, this intensity reduction is caused by holes accumulation in region IV that provokes the dip in the integrated PL intensity described in Figure~\ref{Figure3}~(c).

The effective temperature derived from each spectral tail is plotted in Figure~\ref{Figure5}~(c) for PL1 (green circles) and PL2 (red triangles) emissions. The $I(V)$-characteristic voltage shift has been added in the background. In both cases, $T^{\text{e-h}}_{\text{eff}}$ grows with the applied voltage confirming the carriers heating by the external electric field. The dip of the PL intensity, linked to the GaSb band narrowing, prevents obtaining a $T^{\text{e-h}}_{\text{eff}}$ from the II+III+V-tail as the voltage shift, $\Delta V$, increases. Note, in Figure~\ref{Figure5}~(c), that the $T^{\text{e-h}}_{\text{eff}}$ associated to the I-tail of the PL1 emission (open green circles) can be obtained just for bias voltages between $0.45 \text{ V}$ and $0.80 \text{ V}$. Outside this range, the broadening of the spectra masks this part of the emission band. In turn, the I-tail of the PL2 emission intensifies, widening the detection range (open red triangles) from $V=0$ up to $V=1.30 \text{ V}$. The intensification of the I-tail can be explained by the low laser excitation energy, which favors the photogeneration in the deepest regions, I and II. 

Another striking result in Figure~\ref{Figure5}~(c) is the $T^{\text{e-h}}_{\text{eff}}$ obtained from the II+III+V-tail of the PL1 emission (solid green circles) that was expected to be higher than the PL2 temperatures (solid red triangles), as reported in the preceding work of Vandriel \textit{et al.},~\cite{Vandriel1992} where the effective temperature increases for higher excitation energies. However, the PL1 temperatures for the II+III+V-tail, which are similar to the EL temperatures presented in Figure~\ref{Figure4}~(f), are also lower than the PL2 temperatures. This seeming anomaly will be tackled in the next section by weighting the effects of hot-carriers local cooling and heating in terms of the excitation energy.

\textbf{Thermalizing electrons and holes.} In order to understand these singular results, we proposed a model to account for the dependence of the effective temperature on the carriers dynamics and the experimental conditions, also assessing the thermalization between electrons and holes.~\cite{Shah1985,Kalashnikov1977,Shah1992} When the RTD is subjected to electrical and optical excitations, the averaged carriers energy-loss rate per unit volume in a particular region of the heterostructure can be considered as
\begin{equation} \label{ELR}
    n_i \left\langle \frac{dE_i}{dt} \right\rangle = j_i F_{\text{local}} + n_i \frac{\Delta \mu^{\text{in}}_i}{\tau^{\text{in}}_i} - n_i \frac{\Delta \mu^{\text{out}}_i}{\tau^{\text{out}}_i}
\end{equation}
where $n_i$ ($i=$ e,h) is the carriers density. The first term on the right represents the volumetric electrical power density dissipated by a certain heterostructure region, where $j_i$ is the current density and $F_{\text{local}}$ is a local electric field. The heating/cooling powers, $\Delta \mu^{\text{in/out}}_i / \tau^{\text{in/out}}_i$, are  produced by other sources besides the currents, such as optical excitation or evaporative processes, respectively, with  $\tau_i$, the carriers relaxation time. By substituting eq. \ref{Ei} into eq. \ref{ELR}, the carriers temperature can be expressed as
\begin{equation} \label{Ti1}
    T_i = T_{\text{L}} + \Delta T^{\text{exc}}_i + \frac{\tau^\varepsilon_i}{k_{\text{B}} n_i} j_i F_{\text{local}}
\end{equation}
with $\Delta T^{\text{exc}}_i = \left( \tau^\varepsilon_i / k_{\text{B}} \right) \left[ \left(\Delta \mu^{\text{in}}_i / \tau^{\text{in}}_i \right) - \left(\Delta \mu^{\text{out}}_i / \tau^{\text{out}}_i\right) \right]$ representing the balance in the carriers temperature provoked by carrier excitation or evaporation. This balance also depends on the excitation energy. By increasing the incoming photons' energy, the chances for the local extraction of high energetic carriers can be enhanced, reducing the average excess energy, and lowering the effective temperatures. This explains the seeming paradox of the temperature reduction in the II+III+V band of the PL1 response highlighted previously when discussing Figure~\ref{Figure5}~(c).

\begin{figure}
	\includegraphics[width=\columnwidth]{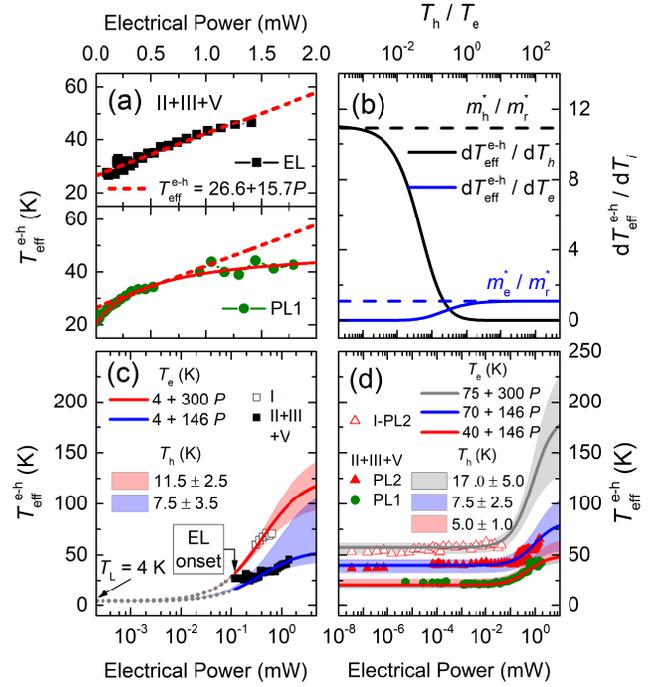}
	\caption{\label{Figure6} (a) $T^{\text{e-h}}_{\text{eff}}$ as a function of the applied electrical power obtained from the GaSb EL (black squares) and PL1 (green dots) spectral tail coming from regions II, III, and V. Red dashed lines correspond to a linear fit extrapolated to zero applied electrical power.  (b) Calculated derivative of $T^{\text{e-h}}_{\text{eff}}$ with respect to $T_i$ for $i=$h (black line), and $i=$e (blue line) as a function of the ratio $T_{\text{h}}/T_{\text{e}}$. Horizontal dashed lines indicate the limits where changes in $T_{\text{h}}/T_{\text{e}}$ produce the greatest variations in $T^{\text{e-h}}_{\text{eff}}$. Simulations of the electrical power dependence for $T^{\text{e-h}}_{\text{eff}}$ carried out for the data extracted from (c) EL and (d) PL spectral tails. Symbols indicate the data obtained from the spectral tails coming from different GaSb regions. The best simulations are displayed as color solid lines.  Color shaded regions represent simulations varying $T_{\text{h}}$. In (c), the mathematical extrapolations to zero electrical power are depicted as gray doted lines. The simulation for the PL1 data (red solid line) in (d) is also plotted in the bottom panel of (a) to show the good agreement between simulation and experiment.}
\end{figure}

The thermalization among electrons and holes can be assessed by analyzing the evolution of the e-h effective temperature on the applied electrical power, $P$, obtained as the product between the measured current and the applied voltage, $P=I V$. The value of $T^{\text{e-h}}_{\text{eff}}$ extracted from the II+III+V-tail of the GaSb emission as a function of the applied electrical power is shown in Figure~\ref{Figure6}~(a). Top (bottom) panel displays the $T^{\text{e-h}}_{\text{eff}}$ obtained from EL (PL1) emissions. As expected, $T^{\text{e-h}}_{\text{eff}}$ grows with electrical power.

The volumetric electrical power density can be correlated with $P$ using the relation $j_i F_{\text{local}} \approx \xi P / \mathcal{V}_{\text{local}}$, being $\xi$ a leverage factor and $\mathcal{V}_{\text{local}}$ the local volume of the considered region. Then, eq.~\ref{Ti1} takes the form
\begin{equation} \label{Ti2}
    T_i = T_{\text{L}} + \Delta T^{\text{exc}}_i + \alpha_i P
\end{equation}
where $\alpha_i = \xi \tau^\varepsilon_i / \left( k_{\text{B}} n_i \mathcal{V}_{\text{local}} \right)$. Hence,  according to eq. \ref{EqTeff}, if carriers thermalize among them during external excitation, then $T^{\text{e-h}}_{\text{eff}}=T_{\text{e}}=T_{\text{h}}$, and a linear dependence with $P$ is also expected for $T^{\text{e-h}}_{\text{eff}}$. The linear fit for $T^{\text{e-h}}_{\text{eff}}$ is represented in Figure~\ref{Figure6}~(a) by the red dashed lines for both the EL (top panel) and PL (bottom panel) effective temperatures extracted from the GaSb spectral tail of the emission produced by the II, III, and V regions.

Considering that during EL there are no external excitation sources besides the applied voltage, then  $\Delta T^{\text{exc}}_i=0$. Although the EL temperatures seem to follow a linear dependence with $P$, the fitted lattice temperature, $T_{\text{L}}=26.6\text{ K}$, lacks any physical sense since in the dark, at zero applied electrical power, the carriers are expected to thermalize with the lattice ($T^{\text{e-h}}_{\text{eff}}=T_{\text{L}}=4\text{ K}$). Moreover, the PL1 temperatures clearly can not be described by a linear dependence with $P$, as displayed in the bottom panel of Figure~\ref{Figure6}~(a). This precludes the assumption of thermalized electrons and holes as anticipated by Bonch-Bruevich and Kalashnikov,~\cite{Kalashnikov1977} who established that contrasting effective masses and scattering times between hot-carriers are linked to differences in their effective temperatures leading to $T_{\text{e}}\neq T_{\text{h}}$.

According to eq.~\ref{EqTeff}, the weights of electron or hole contributions to the e-h effective temperature depend on their effective masses. The relative effect of varying individual temperatures can be evaluated by examining the derivative  
\begin{equation} \label{der}
    \frac{d T^{\text{e-h}}_{\text{eff}}}{dT_{\text{e(h)}}}=\frac{m^*_{\text{e(h)}}}{m^*_{\text{r}}}\left( 1+\frac{m^*_{\text{e(h)}}T_{\text{e(h)}}}{m^*_{\text{h(e)}}T_{\text{h(e)}}} \right)^{-2}
\end{equation}
as a function of the ratio between carriers temperature, $T_{\text{h}}/T_{\text{e}}$, as depicted in Figure~\ref{Figure6}~(b). The derivative with respect to the holes temperature (black solid line) shows that when $T_{\text{h}}/T_{\text{e}} \to \infty$, $T^{\text{e-h}}_{\text{eff}}$ is not affected by changes of $T_{\text{h}}$, but when $T_{\text{h}}/T_{\text{e}} \to 0$, $T^{\text{e-h}}_{\text{eff}}$ is strongly influenced by varying $T_{\text{h}}$. For the derivative with respect to the electrons temperature (blue solid line), the contrary is expected. In both cases, the variation of the carriers effective temperature is limited by the ratios $m^*_{\text{h}}/m^*_{\text{r}}$ (black dashed line) and $m^*_{\text{e}}/m^*_{\text{r}}$ (blue dashed line) for holes and electrons, respectively. This dependence indicates that fluctuations in holes temperature can also be detected by measuring $T^{\text{e-h}}_{\text{eff}}$, if $T_{\text{h}}$ is low compared with $T_{\text{e}}$. That is the reason the cooling process observed in Figure~\ref{Figure4}~(d), ascribed to holes evaporation from the OW, has been observed under illumination. Under such circumstances, the light excitation leads to hotter non-equilibrium electrons arriving at the OW so that the reduction in the local effective temperature of holes is better resolved. 

The difference between carriers temperature is also responsible for the non-linear dependence of $T^{\text{e-h}}_{\text{eff}}$ with the applied electrical power. Using eqs.~\ref{EqTeff} and \ref{Ti2} we calculated $T^{\text{e-h}}_{\text{eff}}$ as a function of $P$, as shown in Figures~\ref{Figure6}~(c) and (d) on a semi-logarithmic scale for the EL and EL+PL temperatures, respectively. Color solid lines represent the best simulations keeping $T_{\text{h}}$ constant, while color shaded bands indicate the possible responses by sweeping the holes temperature around the best value, $T_{\text{h}}$, with a deviation, $\Delta T_{\text{h}}$. 

As discussed before, in the case of the EL excitation we have $T_{\text{L}}=4 \text{ K}$ and $\Delta T^{\text{exc}}_{\text{e}}=0$, then simulations suggest that $\alpha_{\text{e}}$ should be around $300 \text{ K mW}^{-1}$ for the effective temperatures extracted from the I-tail (red), which is almost twice the value of $\alpha_{\text{e}}=146 \text{ K mW}^{-1}$ obtained for the effective temperatures extracted from the II+III+V-tail (blue) as presented in Figure~\ref{Figure6}~(c). In turn, the best holes temperatures should be close to $11.5 \text{ K}$ in region I, and $7.5 \text{ K}$ in regions II, III, and V, whereas electrons can reach temperatures of hundreds of Kelvin at the highest electrical powers. Since holes are created by impact ionization above the EL onset, the extrapolation of the simulations to $P=0$ (gray dotted lines) has just mathematical meaning. 

The simulations for the PL temperatures displayed in Figure~\ref{Figure6}~(d) were performed taken as reference the $\alpha_{\text{e}}$ parameters obtained for $T_{\text{e}}$ using the EL temperatures. The best simulations show a good agreement with the experimental results in the whole range of the applied electrical power, as can be verified in the bottom panel of Figure~\ref{Figure6}~(a), where the best simulation for the PL1 temperatures is displayed as a solid red line.
In accordance with these simulations, holes temperature is expected to be closer to the lattice temperature at regions II, III, and V (red and blue bands), but of the order of tens of Kelvin at region I (gray band), when the heterostructure is illuminated. Extrapolations of the simulations to $P=0$ show that $T_{\text{L}}+\Delta T^{\text{exc}}_{\text{e}}$ should be close to $40 \text{ K}$ for the PL1 response and to $70 \text{ K}$ for the PL2 emission, demonstrating that the low-energy laser is more efficient for heating electrons in the GaSb layers than the high-energy laser, due to its greater penetration into the heterostructure. The simulations confirm that, besides the linear dependence of the carriers temperature with the applied electrical power, the non-thermalization between electrons and holes leads to a non-linear dependence in the case of $T^{\text{e-h}}_{\text{eff}}$.

Following these results, a gradient map of the carriers temperature throughout the heterostructure can be sketched as shown by Figures~\ref{Figure2}~(a) and (b) for EL and EL+PL conditions, respectively. Color-gradient background represents the independent variations of the electrons (holes) temperature, $T_{\text{e}}$ ($T_{\text{h}}$), with $T_{\text{e}}$ higher (red-yellow gradient) than $T_{\text{h}}$ (blue gradient). When a forward bias voltage is applied, carriers are expected to reach the highest temperatures at region I and the OW (dark colors) due to the highly doping profiles, whereas at regions IV and V, electrons and holes attain the lowest temperatures (light colors).

\section{Concluding remarks}

The segmentation of the optical characterization of the semiconductor heterostructure allowed mapping the local thermalization processes through the characterization of the effective temperature of non-equilibrium carriers. The combination of heating and cooling mechanisms builds a non-trivial picture of energy relaxation along the heterostructure. Several factors as the difference between carriers effective masses and the carriers populations, as well as optical and electrical excitations, prevent electrons and holes from thermalizing among them and with the lattice. Then, electron and hole populations should be considered independent systems with independent temperatures at the conduction and valence band. Changes in the effective temperature depend on the relation between the effective masses of the charge carriers and the ratio between the absolute carrier temperatures. If the holes temperature is much lower than the electrons temperature, as suggested by our results, then the holes temperature can produce both, noticeable fluctuations in the effective temperature and the non-linear dependence of the carriers effective temperature with electrical power observed in this work.

\section{Methods}

We study an $n$-type, Te-doped RTD heterostructure grown \textit{via} molecular beam epitaxy on a GaSb(100) substrate as 
described in a previous work.~\cite{Pfenning20171} First, $300 \text{ nm}$ of $n$-type GaSb are grown, as indicated by region I in Figure~\ref{Figure1}(b), with $n=1\times 10^{18} \text{ cm}^{-3}$, followed by $120 \text{ nm}$ $n$-type GaSb (region II) with $n=5\times 10^{17} \text{ cm}^{-3}$. The DBS is formed by a 5-nm thick GaAs$_{0.05}$Sb$_{0.95}$ QW, sandwiched by two 4.5-nm thick AlAs$_{0.08}$Sb$_{0.92}$ barriers and two intrinsic GaSb spacer layers, identified as regions III and IV with thicknesses of $20 \text{ nm}$ and $27 \text{ nm}$, respectively. A 7-nm thick intrinsic GaAs$_{0.05}$Sb$_{0.95}$ layer is deposited after region III during the growth process to form a pseudomorphical prewell. Region V is formed by $120 \text{ nm}$ of $n$-type GaSb with $n=5\times 10^{17} \text{ cm}^{-3}$, followed by an $n$-type 220-nm thick Al$_{0.30}$Ga$_{0.70}$As$_{0.03}$Sb$_{0.97}$ OW, with $n=1\times 10^{18} \text{ cm}^{-3}$. The heterostructre is completed with an $n$-type 10-nm thick GaSb capping layer with $n=1\times 10^{18} \text{ cm}^{-3}$, and a Au/Ti ring contact to allow for optical access, as illustrated in Figure~\ref{Figure1}(a).

To explore the optical response in the dark and under illumination, we studied the EL and PL emissions, respectively, taking advantage of the transport characteristics of the RTD. All the measurements were performed at $4\text{ K}$ in an ultra-low vibration cryostat (Attocube AttoDRY1000), associated with a homemade confocal microscope and a SourceMeter (Keithley 2400). Since the bandgap energies of the GaSb layers and the AlGaAsSb OW are expected to be $0.81 \text{ eV}$ and $1.20 \text{ eV}$, respectively, as indicated by Figure~\ref{Figure1}(b),~\cite{Birner2007} two incident diode lasers were used as excitation sources, one emitting at $\hbar\omega_1=1.88 \text{ eV}$ (Toptica Smart iBeam) to optically excite the entire heterostructure, and another emitting at $\hbar\omega_2=1.15 \text{ eV}$ (PicoQuant LDH) that allows exciting just the GaSb layers and the DBS, avoiding photogeneration of holes at the OW and increasing light penetration. Both lasers were operated in continuous mode at a power density of $4.2\times10^{4}\text{ W }\text{cm}^{-2}$, allowing for the characterization of the influence of the excitation energy on the thermalization processes. The optical responses produced by the lasers, labeled as PL1 and PL2 respectively, were dispersed by a 50-cm spectrometer (Andor Shamrock) and detected by a high-resolution InGaAs diode array (Andor DU491A - 1024 pixels/line).

\section{Acknowledgments}

The authors are grateful for financial support by the BAYLAT and by Brazilian agencies: the Fundação de Amparo à Pesquisa do Estado de São Paulo (FAPESP) - grants No. 2013/18719-1, No. 2014/07375-2, No. 2014/19142-2, No. 2014/02112-3, No. 2015/13771-0 and No. 2018/01914-0, the Conselho Nacional de Desenvolvimento Científico e Tecnológico (CNPq) and the Coordenação de Aperfeiçoamento de Pessoal de Nível Superior (CAPES) - Finance Code 001. This work was also partially funded by the MSCA-ITN-2020 QUANTIMONY from the European Union’s Horizon 2020 programme under Grant agreement ID: 956548.

The revised accepted version can be found at \href{https://pubs.acs.org/doi/10.1021/acs.jpcc.1c02173}{https://pubs.acs.org/doi/10.1021/acs.jpcc.1c02173}

\bibliography{manuscript_JPCC}

\end{document}